\documentclass[aps,pre,
   superscriptadd ress,
   eqsecnum,
   floatfix,
  noshowpacs,
   preprintnumbers,
   nofootinbib
   ]{revtex4-1}

\usepackage{color}
\usepackage{framed}
\usepackage{subfigure}
\usepackage{time}
\usepackage{graphicx}
\usepackage{enumerate}
\usepackage{latexsym}
\usepackage{bm}
\usepackage{upgreek}
\usepackage{amsthm}
\usepackage{amssymb}
\usepackage{mathrsfs}
\usepackage{amsmath}
\usepackage{physics}                           
\newcommand{\B} {\mathcal{B}}
\newcommand{\M}{\mathcal{M}}
\newcommand{\alphap}{\vec{\alpha} \cdot \vec{p} }
\newcommand {\tc}{{\tilde c}}
 \newcommand{\tp}{\tilde{p}}
  \newcommand{\tN}{\tilde{N}}
\newcommand{\be}{\begin{equation}}
\newcommand{\bq}{\begin{equation}}
\newcommand{\ee}{\end{equation}}
\newcommand{\eq}{\end{equation}}
\newcommand{\bea}{\begin{eqnarray}}
\newcommand{\eea}{\end{eqnarray}}
\newcommand{\ba}{\begin{eqnarray}}
\newcommand{\ea}{\end{eqnarray}}

\begin{document}

\preprint{LA-UR- 20-22766}
\title{Neutrino condensation from a New Higgs-like Interaction}

\author{Alan Chodos}
\email{alan.chodos@uta.edu}
\affiliation{Dept. of Physics, University of Texas at Arlington, 502 Yates Street, Box 19059, Arlington, TX 76019 }

\author{Fred Cooper} 
\email{cooper@santafe.edu}
\affiliation{The Santa Fe Institute, 1399 Hyde Park Road, Santa Fe, NM 87501, USA}
\affiliation{Theoretical Division and Center for Nonlinear Studies,
   Los Alamos National Laboratory,
   Los Alamos, NM 87545}

\begin{abstract}
We revisit the scenario of having a new interaction between neutrinos and a Higgs-like  scalar suggested by short-baseline neutrino anomalies. We find that there are two possible attractive
channels in the resulting effective 4-fermi theory which lead to a neutrino condensation in cosmic neutrinos and the creation of a neutrino superfluid at low temperatures and finite chemical potential.  We find that at the minimum of the effective potential $V$ the condensates are mostly made up of pairs of left-left+ right-right composites, with a tiny admixture of left-right + right-left   composites which differs from the scenario of Kapusta which focused on the energetically unfavored   left-right + right-left  scenario.  We obtain an approximate equation for the gap
$  \Delta^2 = 4 m_H (\mu -m) \rm{Exp} [{-\frac{\pi ^2}{G_{eff} \mu  \sqrt{\mu ^2-m^2}}}]$, where $m_H$ is the mass of the Higgs scalar, and $G_{eff}$ is the effective (renormalized) 4- Fermi interaction.  Recent constraints on cosmological neutrinos suggest the possibility that ~$ \rm{ log_{10} } ( \rm{G_{eff} MeV^2}) = -1.72$  which would lead to an exponentially small gap. 
\end{abstract}
\pacs{11.15.Kc,03.70.+k,0570.Ln.,11.10.-z}
\date{\today, \now \ MDT}
\maketitle

\section{Introduction}
It is virtually certain that the cosmic neutrino background (CNB) exists, even though it has yet to be detected, lurking at a slightly lower temperature than its more renowned cousin, the cosmic microwave background. If and when the CNB is amenable to observation, it may well exhibit some interesting dynamical properties. 

One reason so to suppose is the tantalizing numerical coincidence expressed by  $\Lambda M_p^2= m_\nu^4$,  Here $\Lambda$ is the measured cosmological constant, $M_p$  is the Planck length, and, as a plausible assumption, we take the neutrino mass $m_\nu$ to be of the same order of magnitude as the measured neutrino mass differences. Another way of stating this result is that the scale of dark energy density is given by the neutrino mass. 

This may be a pure coincidence, of no further significance than, say, the Koide relation \cite{Koide}, or the fact that the proton to electron mass ratio is $6 \pi^5$ . Nevertheless, late in the last millennium, it prompted Caldi  and one of the present authors to conjecture \cite{Chodos1}  that the cosmological vacuum was home to a neutrino condensate, as a way of seeing why the neutrino mass and the dark energy might be connected. The rough argument was based on assuming an effective 4-neutrino interaction at low energies. If a $<\nu \nu> $condensate formed, then schematically a neutrino mass would be generated by terms of the form $\nu \nu < \nu \nu >$ , and the pure condensate term $ <\nu \nu> < \nu \nu >$  would contribute to the cosmological constant. Here we are being very generic: $< \nu \nu > $ is meant to stand for any type of neutrino-neutrino or neutrino-antineutrino pairing, as dictated by whatever attractive interactions exist that could induce a condensate to form. Of course, even if this scenario is realized, one still has to investigate whether it can lead to the simple numerical relationship mentioned above.

As a first step, the authors of  \cite {Chodos1}  examined the effective 4-neutrino interaction due to the exchange of the $Z$ boson. They looked for pairing of the superfluid neutrino-neutrino type, in the presence of a chemical potential. If there is an attractive channel, a solution to the relevant gap equation is guaranteed, because, in the absence of a gap, the interaction becomes infinite as one approaches the Fermi surface. However, the authors found that no attractive channel exists, perhaps not surprising in view of the observation that vector exchange produces repulsion between like charges.

A few years later, the subject was advanced by the work of Kapusta \cite{Kapusta}, who considered the exchange of the Higgs boson, which indeed produces an attractive channel. As we shall see below, there are in fact two such channels; Kapusta examined the one that couples left-handed to right-handed neutrinos. Once again, if one adds a chemical potential one finds a solution to the gap equation, and hence evidence for a condensate. However, the coupling $g$ of the Higgs boson to neutrinos (assuming that that is how the neutrinos get their mass) is exceedingly small, since the vacuum expectation value of the Higgs is already determined by the Standard Model. From the gap equation, one infers that the size of the condensate depends exponentially on the coupling:   $< \nu \nu > \propto e^{-1/g^2} $  so the magnitude of a condensate generated in this way is unfortunately totally negligible.

Lately, there has been increased interest in various forms of non-standard neutrino interactions \cite{nonstandard} , invoked to address a range of issues. One possibility is the existence of a new, ``neutrinophilic''  Higgs-like boson \cite{Asaadi}. It is hypothesized to couple preferentially to neutrinos, in the same manner as the ordinary Higgs would, but with a much lower value for the vacuum expectation value, thus permitting the coupling constant to be similar in magnitude to the coupling of the ordinary Higgs to the charged leptons. As a result, it could generate a condensate similar to the one found by Kapusta, but with a considerably larger value. 

In fact, recent work \cite{Park, Lancaster}, considers a model of neutrinos with a scalar interaction characterized by an effective coupling 
$G_{eff}=\frac{g^2}{m_H^2}$ whose magnitude, given by ~$ \rm{ log_{10} } ( \rm{G_{eff} MeV^2}) = -1.72$, at least 31 orders of magnitude larger than the coupling in reference \cite{Kapusta}.  (This model has certain advantages over the standard Lambda Cold Dark Matter (LCDM), including a possible resolution of the tension in measurements of the Hubble parameter. Furthermore, in  \cite{Asaadi}, the mass of the neutrinophilic Higgs consistent with the short baseline neutrino anomalies is taken to be of order $3 \times 10^4 ev$.

In the following analysis we explore the possibility of a new Higgs in the simplified situation where there is only one flavor of neutrino. As in past work, we look for a pairing of superfluid type. We note that the relevant Fierz rearrangement allows for condensates with two different sets of quantum numbers, the Lorentz-invariant matrix structure, $ i \gamma^1 \gamma^3$ , which does not flip the handedness of the neutrino, and the structure $\gamma^2 \gamma^5$ , which does. It is the latter that was considered in  \cite {Kapusta} for the case of the ordinary Higgs. In our analysis, we shall find, for a range of the parameters, that the former condensate in fact dominates the dynamics.

As shown in reference \cite{Bhatt} ,  it is also possible to treat the neutrino flavors as Majorana. The introduction of a sterile neutrino allows for one of the neutrinos to obtain a mass via the seesaw mechanism (in their case the seesaw operates between 0.1eV and 0.001eV), and the further introduction of a scalar particle induces condensation, which they take to be of the form considered by Kapusta. They argue that even though lepton number is not conserved in such a model, in a cosmological setting one can still define a chemical potential provided that the expansion rate of the universe exceeds the interaction rate of the particles.

We shall not consider this possibility in what follows, although one can contemplate extending our work in that direction in the future. In this paper the Majorana nature of the neutrino emerges as a result of the condensation (i.e. lepton number is broken spontaneously) and is not introduced a priori. But we do generalize previous work by allowing for simultaneous condensates in two separate attractive channels. 

In section II, we  derive the effective action for the condensate fields  in mean-field approximation, using the well-known Hubbard-Stratonovich procedure. In section III, we  determine the extrema of the effective potential and discuss the effects of coupling constant renormalization.  We also determine the renormalized gap equation and the renormalized effective potential.  In section IV we consider the analytic approximation that the gap equation integral is dominated by the peak of the integrand at the fermi surface and compare the result with the numerical calculation of the gap equation.  We summarize our results in section V. 

\section{Dynamics of Neutrinos when there is interaction with a new Higgs particle}

As described in the introduction, we assume,  for simplicity, one species of neutrino interacting with a Higgs scalar described by the Lagrangian:
\ba
{\cal L}_s &&=  \frac{1}{2} (\partial_u \phi \partial^\mu \phi) - V[\phi], \nonumber \\
V[\phi]&& = -\mu^2 \phi^2/2  + h \phi^4/4,
\ea
with  $\mu^2 > 0$. 
The Higgs potential gives the scalar particle a vacuum expectation value 
\bq
\phi_0=v = \sqrt {\mu^2  /h} ,
\eq
and a tree-level mass 
\bq
m_H^2 = \frac{d^2 V}{d \phi^2} |_{\phi=v} = 2 h v^2. 
\eq
We assume that the neutrino is a Dirac particle having both right and left handed components.  In the Dirac representation

\ba
\psi_L &&= \frac{1}{2} (1- \gamma_5) \psi= \frac{1}{\sqrt 2} \begin{pmatrix}
\nu_L \\
-\nu_L \\
\end{pmatrix} \\ 
\psi_R &&= \frac{1}{2} (1+ \gamma_5) \psi=  \frac{1}{\sqrt 2}\begin{pmatrix}
\nu_R \\
\nu_R \\
\end{pmatrix} , \\
\ea
 where $\nu_R $ and $\nu_L$ are two component spinors so that 

\bq
\psi = \frac{1}{\sqrt 2}\begin{pmatrix}
\nu_R+\nu_L \\
\nu_R-\nu_L \\
\end{pmatrix} .
\eq 

The neutrino Lagrangian including interaction with the Higgs particle is given by 
\bq
{\cal L}_f = \bar{\psi} i \gamma^\mu \partial_\mu \psi - g \bar{\psi} \psi \phi .
\eq
The leading  effect of the Higgs particle is to  give a Dirac mass to the neutrinos:  $m= g v$, where $v= \langle \phi \rangle$, as well as induce an effective four-fermi interaction 
between the neutrinos. 
This leads to a low energy effective neutrino Lagrangian :
\bq \label{Lag}
{\cal L}_f = \frac{i}{2}(  \bar{\psi}  \gamma^\mu \partial_\mu \psi- \partial_\mu \bar \psi \gamma^\mu \psi) - m\bar{\psi} \psi +  \frac {g^2} {m_H^2}( \bar{\psi} \psi )^2 .
\eq
We can write this as 
\bq \label{lag2}
{\cal L}_f = \frac{1}{2} (\psi^\dag A \psi - \psi A^T \psi^\dag )  + \M_{\alpha \beta \gamma \delta}  \psi^\dag_\alpha \psi_\beta \psi^\dag _\gamma \psi_\delta ,
\eq
where choosing the metric diag $ \{1,-1,-1,-1 \} $
\bq
A = i  \partial_0 + i \gamma^0 \gamma^k \partial_k -  \gamma^0 m .
\eq
We would like to know if  the exchange of the Higgs-like particle in the ``t"  channel of the neutrinos can lead to the formation of bosonic condensates in the ``s" channel.  To see whether this is true we first reorganize the 4-Fermi interaction to be in the form of fermion pairs  interacting. This then 
leads to a BCS-like interaction.  In the Hamiltonian operator formalism one would, in the mean field approximation, replace two of the fields in the 4 fermi interaction,  $\psi_i \psi_j$,  with their expectation value, so that  schematically the field operators 
\bq
\psi^\dag \psi^\dag   \psi \psi  \rightarrow \psi^\dag   \psi^\dag   \langle  \psi  \psi  \rangle  = 
\psi^\dag   \psi^\dag  ~ \sigma
\eq
where $\sigma$ is shorthand for a composite bosonic state of spin zero or 1. This approach was taken by Kapusta \cite{Kapusta}. 
  In the path integral formulation of the problem, one equivalently introduces these states using a Hubbard-Stratonovich formalism which converts the quartic  fermionic interaction into an equivalent trilinear interaction between 2 fermions and a boson at the expense of introducing quadratic  bosonic terms into the Lagrangian. One then gets an effective theory in terms of only the bosons by integrating out the now quadratic fermionic degrees of freedom.  Keeping only  the stationary phase contribution to the resulting path integral  over the bosonic degrees of freedom is equivalent  to making  the BCS approximation or mean-field approximation in the Hamiltonian approach.  However in the path integral approach one has a well defined way of obtaining corrections to the mean-field approximation \cite{AF}.  We shall find that the condensate that minimizes the effective potential 
  $V_{eff}$ is quite different from the one found by Kapusta and we also find, by evaluating the gap equation numerically, that what we call the ``drastic approximation" of determining the gap equation from just considering the contribution from the fermi surface is not always justified

We are interested in rearranging the last term in Eq. \eqref{lag2} in order  to understand whether this low energy effective interaction can lead to the usual ``Cooper pairs" ,   $\psi^\dag \psi^\dag$ and $\psi \psi$  found in superfluidity.  To do this one makes a Fierz re-ordering of the 4-Fermi interaction, which means we write
\bq
\M_{\alpha \beta \gamma \delta}= \sum_\lambda \eta_\lambda Q_{\alpha \gamma}^{(\lambda)} Q_{\beta \delta} ^{\star (\lambda)} 
\eq
so that 
\bq
\M_{\alpha \beta \gamma \delta}  \psi^\dag_\alpha \psi_\beta \psi^\dag _\gamma \psi_\delta=- \sum_\lambda \eta_\lambda \psi^\dag_\alpha Q_{\alpha \gamma}^{(\lambda)}\psi^\dag _\gamma  \psi_\beta Q_{\beta \delta} ^{\star (\lambda)} \psi_\delta.
\eq

In the appendix we show that 
\bq
( \bar{\psi} \psi )^2 =- \frac{1}{4} \sum_{\alpha=1}^6  \eta_\alpha (\psi^\dag O^\alpha \psi^\dag )(\psi O^{ \alpha \star} \psi) ,
\eq
Here $\eta_\alpha= \pm 1$.  There are 2 attractive channels, with $\eta_\alpha=  1$ and 4 repulsive ones with $\eta_\alpha=  -1$. 

Since we are interested in neutrino condensation we will focus on the attractive channels. For those cases $O^{ \alpha \star}= - O^\alpha$,
and (ignoring the repulsive channels), 
\bq
( \bar{\psi} \psi )^2  \rightarrow  \frac{1}{4} [(\psi^\dag \sigma^{13}  \psi^\dag )(\psi \sigma^{13}  \psi)+(\psi^\dag \gamma^2 \gamma^5 \psi^\dag )(\psi \gamma^2 \gamma^5 \psi)] .
\eq
so that 
\bq
\M_{\alpha \beta \gamma \delta}  \rightarrow \sum_\lambda Q_{\alpha \gamma} ^{(\lambda)} Q^{\star (\lambda)}_{\beta \delta} ,
\eq
with 
\bq
Q^{(1)}= i \kappa \gamma^1 \gamma^3;, Q^{(2) }=  \kappa \gamma^2 \gamma^5 ,
\eq
and $ \kappa^2 = \frac{g^2}{4 m_H^2}.$  Note that Q is already proportional to $\kappa$.

To make contact with the work of Kapusta, we can write the $\gamma^\mu$ in terms of  the Pauli matrices.  We find that 
\bq
\sigma^{13} = i  \gamma^1 \gamma^3 = - \begin{pmatrix}
\sigma_2, ~0 \\
 \\
0, ~  \sigma_2
\end{pmatrix}, ~~ (\sigma_2)_{ij} = -i \epsilon_{ij}  , 
\ee
so that   
\ba
&& [(\psi^\dag \sigma^{13}  \psi^\dag )(\psi \sigma^{13}  \psi) \rightarrow   \nonumber \\
&&(\nu^\dag_R \sigma_2 \psi^\dag _R +\nu^\dag_L \sigma_2 \nu^\dag_L)(\nu_R \sigma_2 \nu_R + \nu_L \sigma_2 \nu_L) \\
\ea
On the other hand
\bq
\gamma^2 \gamma^5  =   \begin{pmatrix}
\sigma_2, ~~0 \\
 \\
0,  ~-\sigma_2
\end{pmatrix}, 
\ee
so that in  terms of right-handed and left-handed neutrinos,  the attractive channels are:

\ba
&&(\psi^\dag \gamma^2 \gamma^5   \psi^\dag )(\psi \gamma^2 \gamma^5  \psi) \rightarrow   
( \nu^\dag_R \sigma_2 \psi^\dag _L ) (\nu_R \sigma_2 \nu_L ) \nonumber \\
\ea

Kapusta only studies the second possibility where the condensate is a $RL$ composite.  We will find that in our more general framework, this solution is only a relative minimum at an endpoint of the generalized space of solutions. 

We implement the Hubbard-Stratonovich procedure by adding auxiliary fields to the action in such a way as to cancel the terms that are quartic in the neutrino fields.  This does not change the action, since if we perform the  integration over these fields in the path integral we recover the original action. 
Thus we  add to $\mathcal L$ the terms

\bq
-  \frac{1}{\kappa^2} \sum_\lambda \left( B^{(\lambda) \dag} - \kappa Q_{\alpha \gamma} ^{(\lambda)} \psi_\alpha^\dag \psi_\gamma^\dag \right) \left( 
B^{(\lambda) } + \kappa Q_{\beta \delta} ^{\star(\lambda)} \psi_\beta\psi_\delta \right) 
\eq
which then cancels the quartic interaction in Eq.  \eqref{Lag}, to yield:
\bq
{\cal L}_f = \frac{1}{2} (\psi^\dag A \psi  - \psi A^T \psi^\dag )  -  \frac{1}{\kappa^2}\sum_\lambda  B^{(\lambda) \dag}   B^{(\lambda) } + \psi^\dag \cal B\psi^\dag + \psi  \cal B^\dag \psi ,
\eq
where
\ba
{\cal {B}} && = \frac{1}{\kappa} \sum_\lambda  B^{(\lambda)}  Q ^{(\lambda)} ; ~~ {\cal {B} } ^\dag =  - \frac{1}{\kappa}\sum_\lambda  B^{\dag(\lambda)} Q ^{\star (\lambda)}  \nonumber \\
{\cal {B}} && =( B^{(1)}  i \gamma^1 \gamma^3 + B^{(2)}  \gamma^2 \gamma^5 ) ~~ {\cal {B} } ^\dag = ( B^{(1) \dag} i \gamma^1 \gamma^3 + B^{(2)\dag}  \gamma^2 \gamma^5 ), 
\ea
since $Q ^{\star (\lambda)} =- Q ^{(\lambda)}$.

Thinking of $(\psi,\psi^\dag)$ as a column vector $\Psi$ ,  we can represent  ${\cal L}_f $ as 
\bq {\cal L}_f = -  \frac{1}{\kappa^2}\sum_\lambda  B^{(\lambda) \dag}   B^{(\lambda) } + \Psi^\dag S_F^{-1} \Psi ,
\eq
and letting $\psi = \chi + \frac{1}{2} (\B^\dag)^{-1} A^T \psi^\dag$, one obtains

\bq
\int d  B ^\dag d B  \exp[ i \Gamma_{eff} (B^\dag, B) ] ,
\eq
where
\bq 
\Gamma_{eff} = -   \int d^4 x  \left(\frac{1}{\kappa^2} \sum_\lambda B^{\dag (\lambda)} B(\lambda)  + \frac{i}{2} {\rm Tr ~log} \left[ 1+4 A^{-1} \B (A^T)^{-1} \B^\dag \right] \right). 
\eq
and the neutrino inverse propagator at finite density,  with chemical potential $\mu$ is 
\bq
A(x-y)  =( i  \partial_0 + i \gamma^0 \gamma^k \partial_k -  \gamma^0 m - \mu) \delta  (x-y) 
\eq 
This action contains all the dynamics of the bosonic degrees of freedom.  The quadratic term is a bare mass term for the bosons and the ${\rm Tr ~log}$ term contains all the fermion loop corrections  to the action. It also generates the kinetic energy for the bosons so that they propagate.
The stationary phase approximation to the path integral over the bosonic degrees of freedom is equivalent to the usual BCS approximation used in superconductivity \cite{BCS}. The stationary 
phase point of the action yields the gap equation which determines the  (non-zero) expectation value of the condensate field. 
We will use the Dirac representation of the $\gamma$ matrices with 
\bq
 i \gamma^0 \gamma^k = \sigma^{0k} = i \alpha^k ,
 \eq
 where
 \bq
 \alpha^k = 
\left(
\begin{array}{cc}
 0  & \sigma^k     \\
 \sigma^k  &   0   \\   
\end{array}
\right) .
\eq

In Fourier space we have
\ba
A(x,y) &&= \int \frac{d^4p}{(2 \pi)^4} \left[(p_0- \mu) -  \vec{\alpha} \cdot \vec{p}   - m \gamma^0 \right] e^{-i p \cdot (x-y) } \nonumber \\
&&= \int \frac{d^4p}{(2 \pi)^4} e^{-i p \cdot (x-y) } A[p].
\ea
Here $p^2= p_0^2 - \vec{p} \cdot \vec{p}$.
To correctly incorporate the chemical potential $\mu$ one needs to introduce an $i \epsilon$ prescription, so that 

\bq
A^{-1} (x,y) = \int \frac{d^4p}{(2 \pi)^4} \frac{ \left[ (p_0- \mu ) + \vec{\alpha} \cdot \vec{p}  +m \gamma^0 \right] } {[p_0-\mu +i \epsilon {\rm{sgn} }p_0]^2 -  \vec{p} \cdot \vec{p} -m^2} e^{-i p \cdot (x-y) }
\eq
or
\bq
A^{-1} (x,y)=\int \frac{d^4p}{(2 \pi)^4} \left[(\tilde{p}_0- \mu) -  \vec{\alpha} \cdot \vec{p}   - m \gamma^0 \right]^{-1}e^{-i p \cdot (x-y) }
\eq
where  \bq \label{tildep0}
\tp_0= p_0+i \epsilon {\rm{sgn} }p_0.
\eq

Now consider the quantity 
\bq
X= 4 A^{-1} \B  (A^T )^{-1} \B^\dag .
\eq
which contains two neutrino propagators and two vertices.  This is similar to the fermion loop diagram of the Nambu Jona -Lasinio model \cite{NJL}.  Here we are not so much interested in the dynamics of the condensate field but what are the degrees of freedom of the condensate and what is the nature of the gap equation and the
energy landscape of the condensate space. For this we only need to study the effective potential $V_{eff}$
for which  $B^{(i)}$  and the $B^{(i) \dag} $  are constants. The  minima of $V_{eff}$  determine the allowed vacuum states of the theory.  When the fields are constant, $X[p] $ is the integrand of the  vacuum polarization graph at zero momentum transfer. In momentum space we have 
\ba
X[p]&&= - 4 \left[ \frac{1} {\left[( \tilde{p}_0- \mu) -  \vec{\alpha} \cdot (\vec{p} )   - m \gamma^0 \right]}  ( B^{(1)}  i \gamma^1 \gamma^3 + B^{(2)}  \gamma^2 \gamma^5 )  \times \right.   \nonumber \\
&& \left.  \frac{1}{\left[(\tilde{p}_0+\mu) -  \vec{\alpha}^T \cdot \vec{p}   +m \gamma^0 \right]} ( B^{(1) \dag} i \gamma^1 \gamma^3 + B^{(2)\dag}  \gamma^2 \gamma^5 )
\right ]
\ea

Using the (anti-)commutation relations for the $\gamma$  matrices, and performing the matrix algebra, we find that 

\ba
X[p] &&= -4\ \frac{1} {\left[( \tilde{p}_0- \mu) -  \vec{\alpha} \cdot \vec{p}   - m \gamma^0 \right]} \times \nonumber  \\
&& \left[  \frac{1}{\left[(\tilde{p}_0+\mu) -  \vec{\alpha} \cdot \vec{p}   +m \gamma^0 \right]} M_2+  \frac{1}{\left[(\tilde{p}_0+\mu) + \vec{\alpha} \cdot \vec{p}   +m \gamma^0 \right]} M_1 \right] ,
\ea
where
\ba
M_1
&&= B^{(1)}B^{(1) \dag} \bf{1} -\gamma^0 B^{(1)}  B^{(2) \dag} ,
\ea
and 
\ba
M_2
&& = B^{(2)}B^{(2) \dag} \bf{1} - \gamma^0 B^{(2)} B^{(1) \dag}.
\ea
As we shall see below $X[p]$ plays a crucial role in determining the effective potential. 
\section{Effective Potential}
To obtain the effective potential we take the $B^{(i)}$ to be constant , and from 
\bq 
\Gamma_{eff} = - V_{eff} \int d^4 x ,
\eq
we have
\bq
 V_{eff} =(\frac{1}{\kappa^2} \sum_\lambda B^{\dag (\lambda)} B(\lambda) + \frac{i}{2}\int \frac{d^4p}{(2 \pi)^4}  {\rm Tr ~log} \left[ 1+X[p] \right] .
\eq 
This naive effective potential has both quadratic divergences and logarithmic divergences. 
   However, since this is only an effective theory, valid up to the mass of the Higgs particle, we will instead think of this theory having an effective cutoff $\Lambda  \approx m_H$.  It will, however be useful to define a ``renormalized'' coupling constant $\kappa_R^2$ so that the 
   theory only logarithmically depends on the cutoff. 
It is convenient to parametrize the condensates as follows:
\bq
 B^{(1)}=R\cos \theta e^{i \phi_1};~~  B^{(2)}=R\sin \theta e^{i \phi_2},
 \eq
 where we take $0 \leq \theta \leq \pi/2$.
 Then we obtain, letting $\phi= \phi_1-\phi_2$
 \ba
M_1&=&  
 R^2 \left( \cos^2 \theta  \bold{1} -  \frac{\gamma^0 }{2}  \sin {2 \theta}  e^{i  \phi} \right) \nonumber \\
M_2&=&  
 R^2  \left( \sin^2 \theta  \bold{1} -  \frac{\gamma^0 }{2}  \sin {2 \theta} e^{-i  \phi} \right)  \nonumber \\
\ea
 When $\theta = 0$,  $M_1=R^2 \bf{1}, M_2=0$, whereas when $\theta = \pi/2, M_1=0, M_2 = R^2 \bf{1}$.
 We notice that when $\sin 2 \theta=0$,  $M_1,M_2$ are independent of $\phi$.  This occurs at the special cases
 $ \theta=0, M_1=R^2$ and $\theta= \pi/2, M_2 = R^2$.

We can write 
 \bq
 X[p] =A^{-1}[p] Z[p] 
 \eq
 with
 \bq
 A[p]= \tp_0- \mu - \alphap - m \gamma^0,
 \eq
whence 
\bq  V_{eff} =(\frac{1}{\kappa^2} R^2  + \frac{i}{2}\int \frac{d^4p}{(2 \pi)^4}  {\rm Tr ~log} \left[ 1+ A^{-1}  Z[p] \right].
\eq

Taking the derivative  with respect to $\theta, \phi$ we obtain for the stationary points of $V_{eff}$ the conditions:
\ba \label{conditions}
\frac {\partial V_{eff}}{\partial \theta}  &&= \frac{i}{2}\int \frac{d^4p}{(2 \pi)^4}  {\rm Tr} [ (A+Z)^{-1} \frac{\partial Z}{\partial \theta} ]=0 ,  \nonumber \\
\frac {\partial V_{eff}}{\partial \phi}  &&= \frac{i}{2}\int \frac{d^4p}{(2 \pi)^4}  {\rm Tr} [ (A+Z)^{-1} \frac{\partial Z}{\partial \phi } ]=0 . \nonumber \\
%
\frac {\partial V_{eff}}{\partial R^2} &&= \frac{1} { \kappa^2}-  \frac{1} { 4 \pi i R^2}   \int \frac{d^4p}{(2 \pi)^3}  {\rm Tr} [ (A+Z)^{-1} Z =0 .
\ea
One can write
\bq
(A+Z)^{-1} = \frac{N_0}{D_0}; 
 Z=\frac{N}{D}.
\eq
The denominator of the integrand in all the first derivatives in  \eqref{conditions} is the quantity 
\bq \label{dd0} 
 DD_0 = (\tp_0-\lambda_+) (\tp_0-\lambda_{-}) (
 \tp_0+\lambda_+) (\tp_0+\lambda_{-}) ,
 \eq
 where
 \bq
 \lambda_\pm =  (\beta \pm 2 \sqrt {\gamma})^{1/2} ,
 \eq 
 and
 \ba
 \beta &&= \mu^2 + \omega_p^2 + 4 R^2; \, \omega_p^2 = p^2 + m^2; \nonumber \\
 \gamma&&= p^2 (\mu^2 + 4 R^2 \sin^2 \theta) + ( \mu m - 2 R^2 \sin2 \theta \cos \phi)^2 \nonumber \\
 && =  (\omega_p^2-m^2 ) (\mu^2 + 2 R^2( 1- \cos 2 \theta)) + ( \mu m - 2 R^2 \sin2 \theta \cos \phi)^2 .
 \ea
 This will enable us in what follows to perform the $p_0$ integral by contour integration.

The equation  which extremizes  the potential with respect to $\phi$ can be written as
  \ba
\frac {\partial V_{eff}}{\partial \phi}  &&=   -  8    \int \frac{dp_0} {2\pi i}  \frac{d^3p}{(2 \pi)^3} \left[ \frac{ \tc_{0 \phi} }{ 4  DD_0} ,  \right]=0,
 \ea
where  $DD_0$ is given by \eqref{dd0} and 
 \ba 
 && \tc_{0 \phi}  = 8 R^2  \sin 2 \theta \sin \phi  f[p, p_0,\mu, R^2] \nonumber \\
 && f[p, p_0,\mu, R^2]= - m ~\mu   \nonumber  \\
&& -\frac{2 \sin (2 \theta ) \cos (\phi ) \left(m^2 \left(p^2-R^2\right)+p^4-p^2
   (\mu +{p_0}) ^2+R^2 (\mu +{p_0}) ^2\right)}  {m^2+p^2-(\mu +{p_0})^2}.
\ea
Hence we can extremize the potential with respect to $\phi$ by choosing $\sin \phi =0$, $\cos \phi=\pm 1$.
We will restrict ourselves to the case $\cos \phi = \pm 1$ in what follows. 
With that assumption, we can rewrite  $\gamma$ as 
\bq
 \gamma= p^2 (\mu^2 + 4 R^2 \sin^2 \theta) + ( \mu m \cos \phi - 2 R^2 \sin2 \theta )^2 .
 \eq
 We notice that the potential depends on the product of  sign $\mu$ and $ \cos \phi$  which can be $ \pm 1$. 
Writing 
\bq
 \mu \cos \phi= | \mu | \eta = \rho
 \eq
 with $\eta = \pm 1$,  we have
 \bq
  \gamma= p^2 (\mu^2 + 4 R^2 \sin^2 \theta) + ( m \rho    - 2 R^2 \sin2 \theta )^2 .
\eq
We consider the two separate cases $\eta = \pm 1$.  We will find when we study  $V_\pm [R^2, \theta]$  that apart from
the endpoints $\theta = 0, \pi/2$ , where $V_+=V_{-}$, the potential with $\eta=-1$ always has higher energy than the potential with $\eta =+1$. We can treat both cases together by using the parameter $\rho = | \mu | \eta$. We have

 \ba
\frac {\partial V_{eff}}{\partial R^2}  &&=  \frac{1}{\kappa^2} -   8    \int \frac{dp_0} {2\pi i}  \frac{d^3p}{(2 \pi)^3} \left[ \frac{ c_0}{ 4 R^2 DD_0}   \right],
 \ea
with $DD_0$  given by \eqref{dd0} and now 
 \ba
 \frac{c_0}{4 R^2} &&= -  \tN(\tp_0,p)  \equiv N(p^2) - \tp_0^2
 \nonumber \\
&&=\mu ^2+m^2+2 \rho  m \sin 2 \theta +p^2 \cos 2 \theta
   -{\tp_0}^2+4 R^2 \cos ^2 2 \theta.
   \ea
 Doing the $p_0$ integral by contour  integration and
closing the contour in the upper half plane we obtain 

  \ba \label{key}
\frac {\partial V_{eff}}{\partial R^2}  &&=  \frac{1}{\kappa^2}  
- 8   \int   \frac{d^3p}{(2 \pi)^3} \left[ \frac{ \tN (\lambda_+,p) } {2 \lambda_+(\lambda_+^2- \lambda_{-}^2)} -  \frac{ \tN(\lambda_{-} ,p) } {2 \lambda_{-} (\lambda_+^2- \lambda_{-}^2)} \right]. \nonumber \\
&&= \frac{1}{\kappa^2}  
-       \frac{2 } {\pi^2}  \int_0^{p_{max}} p^2 dp   \left[ \frac{ N (p^2) } {  \lambda_+ \lambda_{-}[\lambda_+ +  \lambda_{-}]} + \frac{ 1} {\lambda_+ +\lambda_{-}} \right] \nonumber  \\
&&\equiv  \frac{1}{\kappa^2}- \frac{1}{\pi^2} \int_0^{p_{max} } dp  {\mathcal{I}} (p, R^2, \theta, q_i) .
\ea
where we have used  $\tN (\lambda_\pm , p)  = \lambda_\pm^2 - N(p^2)$ .
Note that $\frac {\partial V_{eff}}{\partial R^2} $ is {\em independent} of $\theta$ when $R^2=0$. 
We  can define a  renormalized coupling constant $1/\kappa_R^2$ as the value of $\frac {\partial V_{eff}}{\partial R^2} $ when $\mu = R^2=0$
and $\theta =0$.
\ba
\frac{1}{\kappa_R^2} &&= \frac{1}{\kappa^2}- \frac{1}{\pi^2} \int_0^{p_{max} } dp  {\mathcal{I}} (p, R^2=0,, \mu=0,m)  \nonumber \\
&&= \frac{1}{\kappa^2}-  \Sigma(H, m). 
\ea
Here $H= p_{max} = m_H$.
Explicitly we have
\bq
\Sigma[H,m]= \frac{2 \left(\frac{1}{2} H \sqrt{H^2+m^2}-\frac{1}{2} m^2 \tanh
   ^{-1}\left(\frac{H}{\sqrt{H^2+m^2}}\right)\right)}{\pi ^2}
   \eq
   For large $m_H/m$ we have that 
   \bq
   \Sigma= \frac{{m_H}^2}{\pi^2}
   \eq
   
  The relations between $\kappa$ and $\kappa_R$ are given by
  \bq
        \kappa_R^2 = \frac {\kappa^2} {1-\kappa^2  \Sigma( H, m)};~~   \kappa^2 = \frac {\kappa_R^2} {1+\kappa_R^2  \Sigma( H, m)}.
    \eq
%
From its definition, we know  $V_{eff} (R^2=0) =0$. Integrating with respect to $R^2$ we obtain
\bq
V_{eff}=\frac{R^2}{\kappa^2}  -  \frac{1}{\pi^2} \int _0 ^{R^2}   dR'^2  \int_0^{p_{max} }  dp ~ {\mathcal{I}} (p, R'^2, \theta, q_i) .
\eq
In terms of the renormalized coupling constant:
\bq \label{renormalizedV}
V_{eff}[R^2, \theta,m,\mu] ] =\frac{R^2}{\kappa_R^2}  -  \frac{1}{\pi^2} \int _0 ^{R^2}   dR'^2  \int_0^{p_{max} }  dp ~ \left[{\mathcal{I}} (p, R'^2, \theta, q_i) -{\mathcal{I}} (p, R'^2=0, \theta=0, \mu=0,m)\right] .
\eq
The gap equation is obtained from the place in $R^2$ where the potential is a minimum so that 
\bq
 \frac{1}{\kappa^2}=\frac{1}{\pi^2} \int_0^{p_{max} } dp ~ {\mathcal{I}} (p, R^2=\Delta^2/4, \theta, m,\mu) .
\eq
Renormalizing the coupling constant we obtain the  renormalized gap equation,
\bq
 \frac{1}{\kappa_R^2}=\frac{1}{\pi^2} \int_0^{p_{max} } dp ~ {\mathcal{I}} _{sub} (p, R^2=\Delta^2/4, \theta, m,\mu) ,
\eq
where we use the subtracted integrand. 
For fixed values of $\theta, m,\mu$ this equation gives the relation between the gap  $\Delta$ and the inverse the renormalized coupling constant $\kappa_R^2$.

 We want the value of  $\theta$ that gives the deepest potential.
When $\sin \phi =0$ ,  one finds
  \ba
\frac {\partial V_{eff}}{\partial \theta } &&=   -  8    \int \frac{dp_0} {2\pi i}  \frac{d^3p}{(2 \pi)^3} \left[ \frac{ c_{0 \theta }}{ 4  DD_0}   \right]
 \ea
where
\bq
 c_{0 \theta }  = 4 R^2 \left(4 \rho m \cos (2 \theta)-2 p^2 \sin (2 \theta )-8 R^2 \sin (2
   \theta ) \cos (2 \theta )\right).
   \eq
   \ba \label{dvdthet}
\frac {\partial V_{eff}}{\partial \theta }  
&&= 2 \int   \frac{d^3p}{(2 \pi)^3} \left[ \frac{ c_{0 \theta }} {2 \lambda_+(\lambda_+^2- \lambda_{-}^2)} -  \frac{ c_{0 \theta }}{2 \lambda_{-} (\lambda_+^2- \lambda_{-}^2)} \right] \nonumber \\
&& = - \frac{1}{2 \pi^2} \int_0^ {p_{max}} p^2 dp    \frac{ c_{0 \theta } } { \lambda_+ \lambda_{-} (\lambda_{+}
 + \lambda_{-}) }.
 \ea
At the stationary points 
    \bq  \label{thetastar} 
    0 = - \frac{1}{2 \pi^2}  \int_0^ {p_{max}} p^2 dp \frac{ 4 R^2 \left(4 \rho  m \cos (2 \theta )  )-2 p^2 \sin (2 \theta )-8 R^2 \sin (2
   \theta ) \cos (2 \theta )\right) }{ \lambda_+ \lambda_{-} (\lambda_+ + \lambda_{-})}.
   \eq
   
\begin{figure} 
   \includegraphics[width=3.4in ]{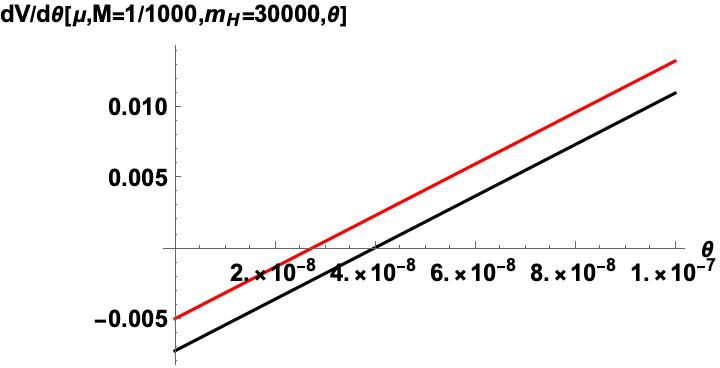}
   \caption{\label{dvdthetmupl}  $\frac{\partial V}{\partial \theta} $  as a function of $\theta $ for  $m=1,M=1/1000, m_H=30000,$ , and $\mu=1.1,1.5, 2$ , $\eta=+1$ }
\end{figure}

   We observe from eqs. \eqref{key}  and  \eqref{dvdthet} that the denominator of each of the integrands contains a factor of $\lambda_{-} $ From the expressions for $\beta$ and $\gamma$ we see that, when  $R^2=0$, $\lambda_{-}$ vanishes at the point $\omega_p= | \mu | $  (this is the Fermi surface), and therefore the integral has a logarithmic singularity as $ R^2 \rightarrow 0$. It is this fact that guarantees a solution to the $R^2$ gap equation no matter how small $\kappa^2 $ may be, provided $| \mu | > m$. 
   One might also conjecture that the integral in eq. \eqref{thetastar}  is dominated by the region around $\omega_p= |\mu| $ , in which case one would conclude that 
\bq \label{Kap} 
    \tan 2 \theta^\star \approx \frac{ 2 \eta |\mu| m} {|\mu^2-m^2|} .
    \eq
However, the term in the numerator proportional to $p^2$  produces a quadratic divergence in the integral as the cutoff tends to infinity; hence we might expect that the integrand is dominated not by the region around $\omega_p= |\mu| $ but rather by $ p^2$ that is some fraction   of  the cutoff  $m_H^2. $.  In that case we expect
\be \label{thetstar3}
    \tan 2 \theta^\star \propto \frac{   |\mu| m} { m_H^2} .
    \eq
This latter result is closer to the numerical findings.  Thus we will find for realistic values of $m_H$  suggested by short baseline experiments,
(i.e $m_H/m_\nu  \approx 3 \times 10^5$) , the value of $\theta^\star$ is close to zero so we obtain a {\em different} condensate than the one suggested by Kapusta. 
Keeping  $M$ fixed ($M=1/1000$) and increasing $\mu$ the minimum in $\theta$ occurs at larger and large $\theta$  and goes like Eq. \ref{thetstar3}, where the red curve corresponds to $\mu = 1.1$ and the black curve to $\mu=1.5$.
This is shown in Fig. \ref{dvdthetmupl} .
Note that the value of $\theta$ that extremizes  the potential does \emph{not} depend on the coupling parameter $1/\kappa^2$.

\begin{figure}[htp]
\begin{center}
\includegraphics[height=.17\textheight, angle =0]{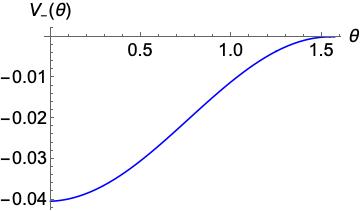}
\includegraphics[height=.17\textheight, angle =0]{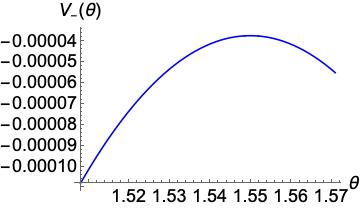}
   \caption{\label{Vminustheta}  $V_{-} ( \theta)  $  as a function of $\theta $ for  $m=1 , \mu=1.5, R^2=1/1000, m_H=20,$ for $\eta=-1$ .
  The figure at the right shows details near the maximum.}
   \end{center} 
   \end{figure}

When $\eta=-1$, the stationary point of the potential in $\theta$ occurs a little below $\theta= \pi/2$.  However it is now a maximum. There is a relative minimum at  the endpoint $\theta =\pi/2$.  This is the state that Kapusta studied. The maximum and the relative minima at   $\theta =\pi/2$ is shown in the rhs of  Fig. \ref{Vminustheta}. There is also a relative endpoint minimum for $\eta=-1$ at $\theta=0$.

\begin{figure}[htp]
\begin{center}
   \includegraphics[height=.17\textheight, angle =0] {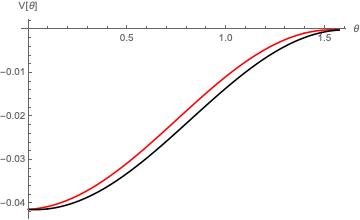}
\includegraphics[height=.17\textheight, angle =0]   {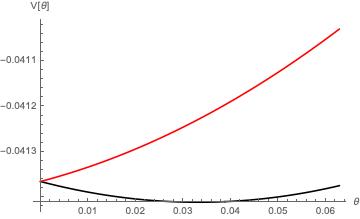}
 \caption{\label{Vcompareksq1}  $ V(\theta, R^2=1/1000)  $  as a function of $\theta $ for$ \mu=1.5, m=1, m_H=20, $ , and $\eta=1 (blue),~ \eta= -1(red) $. }
 \end{center} 
\end{figure}

 Both these endpoint minima are higher than the true minimum for $\eta=+1$ at $\theta^*$.   We display  an example  of this behavior in Fig. (\ref{Vcompareksq1}) where we plot  potentials for $\eta= \pm 1$  as a function of 
  $\theta$ for $\kappa^2=1, m=1, \mu=1.5, m_H=20$. Here we keep $R^2= \Delta^2/4$ fixed to be at  the value $R^2=1/1000$
 We see that the red curve for $\eta=-1$ is above the blue curve for $\eta=+1$ and
  that the blue curve displays the minimum near $\theta=\theta^* \approx .03$. 
  
\subsection{Renormalizing the coupling constant}
\begin{figure}[htp]
\begin{center}
\includegraphics[height=.17\textheight, angle =0]{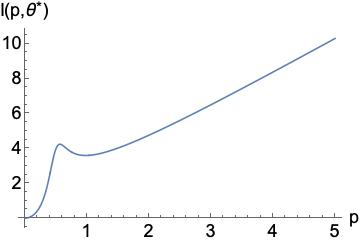}
\includegraphics[height=.17\textheight, angle =0]{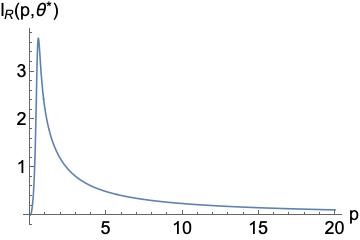}
   \caption{\label{Ip}  $ {\mathcal{I}} (p, R^2=1/1000, \theta=.02, m=1, \mu=1.1) $  as a function of $p$.  The curve on the left is unrenormalized.
  The figure at the right shows the effects of renormalization.}
   \end{center} 
   \end{figure}

\begin{figure}[htp]
\begin{center}
\includegraphics[height=.17\textheight, angle =0]{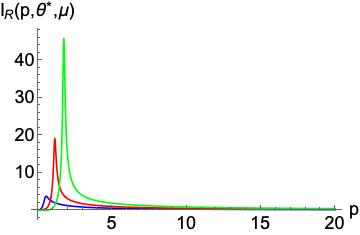}
   \caption{\label{Imu}  $ {\mathcal{I}} (p, R^2=1/1000, \theta^\star) $  as a function of $p$ for increasing values of $\mu$  Blue, Red and Green curves correspond to $\mu= 1.1~,1.5,~2.0$}
   \end{center} 
   \end{figure}

If we use the un-renormalized coupling constant $\kappa^2$ to parametrize the theory,  the integrand in $V[R^2, \theta] $ ,  has a small peak near the Fermi surface,  $\omega_p= |\mu|$   but is larger near the cutoff. Thus one cannot approximate the integrand by the pole contribution.
 However, if we use the renormalized coupling, the integrand now gets a large contribution from the peak at the Fermi surface, as one would anticipate on physical grounds but still has a large tail.  We will compare below the pole contribution approximation in the latter case with the numerical evaluation of  the gap equation.   
The unrenormalized integrand  $ {\mathcal{I}} (p, R^2=\Delta^2/4, \theta^\star, m,\mu) $ has a peak at $p^2+m^2= \mu^2$.  For $m=1, \mu=1.1, m_H=20$ we have that $\theta^\star=.02$ so that  $ {\mathcal{I}} $ has the behavior shown in Fig.  \ref{Ip}. 
This curve shows that if we are going to just keep the pole approximation we needs to use the renormalized coupling constant.  If we keep $m=1, R^2=1/1000$ and vary $\mu$ the spike in the integrand get higher and higher.
This is seen in Fig. \ref{Imu} 

The result that the potential with $\eta=-1$ is always above the potential with $\eta=+1$, as shown in the example of Fig. \ref{Vcompareksq1}  is general so we do not
need to further consider the solution with $\eta=-1$.   

\subsection{renormalized gap equation and renormalized effective potential}
At $\theta^\star$ the renormalized gap equation becomes:
\ba \label{gaptheta}
 \frac{1}{\kappa_R^2}&&= \frac{1}{\pi^2} \int_0^{p_{max} } dp ~ {\mathcal{I}} _{sub} (p, R^2=\Delta^2/4, \theta^\star(m,\mu) , m,\mu) , \nonumber \\
 && \equiv  \frac{1}{\pi^2} g(m, \mu, R^2= \Delta^2/4) 
\ea
This equation relates the gap $\Delta^2$ to the renormalized coupling constant and the parameters $m, \mu$
An analysis of ${\mathcal{I}} _{sub} (p, R^2=\Delta^2/4, \theta^\star(m,\mu) , m,\mu) $  for small values of $R^2$ and 
$\theta^\star$ shows that this integral very weakly dependent on $\theta^\star$, and the value of the renormalized
coupling constant as a function of the gap $\Delta^2$ and the parameters $m, \mu$ is extremely well approximated
by setting  $\theta^\star$ = 0. If we do that we get the simpler equation for the renormalized gap equation:

\ba
\frac{1}{\kappa_R^2}&&=    \int_0^{m_H}   p^2~  dp  \left[ \frac{1}{\sqrt{\Delta^2+
 (\mu+\omega_p)^2} } + \frac{1}{\sqrt{\Delta^2+
 (\mu-\omega_p)^2} } - \frac{2}{\omega_p}   \right] . \nonumber \\
\ea

This integral can be used to define a function of four variables:
\bq
f(\mu, m. m_H, \Delta^2) = \int_0^{m_H}   p^2~  dp  \left[ \frac{1}{\sqrt{\Delta^2+
 (\mu+\omega_p)^2} } + \frac{1}{\sqrt{\Delta^2+
 (\mu-\omega_p)^2} }  -\frac{2} { \omega_p} \right] ,
 \eq
 so that the value of the coupling that allows a solution of the gap equation for given value of $\mu,m,m_H, M= R^2 =\Delta^2 /4$ is given by  
 \bq
 \kappa_R^2 = \frac{\pi^2} { f(\mu,m,m_H ,\Delta^2)} .
 \eq
 
\begin{figure} 
   \includegraphics[width=3.4in ]{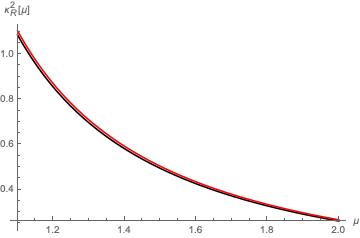} 
   \caption{\label{ksqcompare}  $\kappa_R^2$  as a function of $\mu$   for$M=1/1000, m=1, m_H=20 $.at $\theta=0$ (red) , and at  $\theta=\theta^*(m,\mu)$ (black)   }
\end{figure}

\begin{figure} 
   \includegraphics[width=3.4in ]{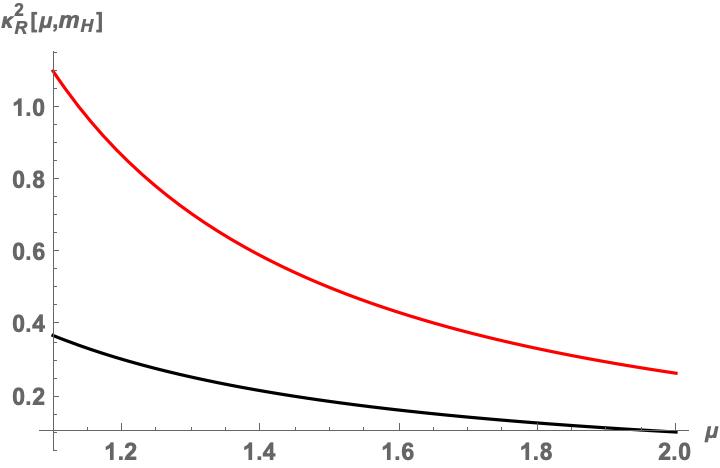} 
   \caption{\label{ksqcompare2}  $\kappa_R^2$  as a function of $\mu$   for$M=1/1000, m=1, m_H=20 $.(red) , and at  $m_H=30000$ (black)   }
\end{figure}

 Note that this determines the effective 4-Fermi interaction $G_{eff}$ in terms of the gap $\Delta^2$.  For what follows, for illustrative
 purposes we will keep $\Delta^2 = 1/1000$  so that $\Delta$ is only slightly smaller than $m$, so that we can see the effect of changing the other parameters visually.  First we want to show that even when we choose the value  $m_H = 20$ very little error is made in letting $\theta^\star=0$.
 We show the slight difference between the  gap equation at $\theta^*$ and $\theta=0$ in Fig.  \ref{ksqcompare} for $m_H=20$.
When $m_H$ increases $\theta^\star$ goes to zero as $1/m^2{_H}$ so this difference rapidly vanishes.  At the nominal value for $m_H$ that comes from the short baseline experiments, the value of $\kappa_R^2$ is lowered by a factor of three as shwon in 
Fig.  \ref{ksqcompare2}  

\begin{figure}[htp]
\begin{center}
\includegraphics[height=.17\textheight, angle =0]{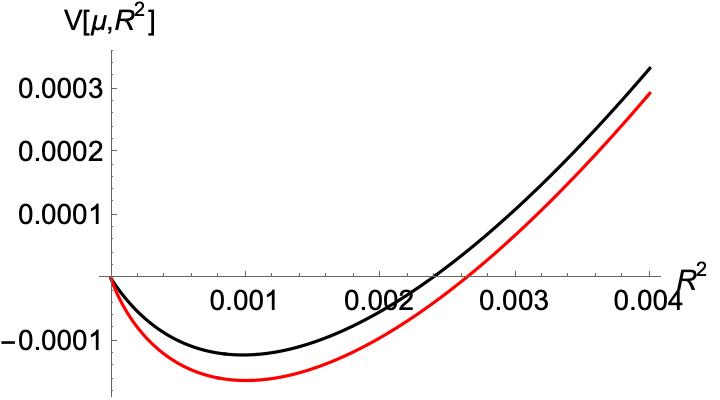}
   \caption{\label{Vrmu}  $V_R[R^2] $  for different  $\mu$ for  $ \Delta^2/4 = 1/1000, m=1, m_H = 3 \times 10^4$ . The colors black and red  correspond to $\mu= 1.1, 1.5$ }
   \end{center} 
   \end{figure}

The equations we have determine the effective potential as a function of the renormalized coupling constant $\kappa_R^2$ which is the effective 4-fermi interaction $G_{eff}$ as well as the parameters $\mu$, $m$, $m_H$ and the gap  $\Delta^2$.  Keeping $m=1$ we are evaluating all masses in terms of the neutrino mass which we assume is around 1/2 eV. 
 Evaluating the effective 
potential $V_{eff}[R^2,m,\mu]$ given in
Eq. (\ref{renormalizedV})  for $\Delta^2=1/1000$ and allowing
 $\kappa_R^2$ to change slightly  with $\mu$ to keep
the gap fixed  one finds that the renormalized effective potential
  $V_{eff}[R^2]$ for $m=1, \mu$ as a function of $R^2$,  is shown in Fig. \ref{Vrmu}.
We see increasing $\mu$ deepens the potential at the minimum.   The minimum is at the $R^2 = \Delta^2$ which we have chosen to be 
$1/1000$ for illustrative purposes.  What we will find below is that  when we obtain an approximate analytical value of $ \Delta^2$, experiments
constrain $ \Delta^2$ to be exponentially small. Nevertheless, $V[R]$ always has a non-zero minimum.

\section{Approximate Analytic Calculation of the Gap Equation } 
Going back to the approximate gap equation at $\theta =0$, we have
\bq
f(\mu, m. m_H, \Delta^2) = \int_0^{m_H}   p^2~  dp  \left[ \frac{1}{\sqrt{\Delta^2+
 (\mu+\omega_p)^2} } + \frac{1}{\sqrt{\Delta^2+
 (\mu-\omega_p)^2} }  -\frac{2} { \omega_p} \right] ,
 \eq

 The integral gets a large contribution near the Fermi surfaces $\omega_p= \pm \mu$. Let us consider the case where  $\mu >0$ so the large contribution to the integral will be from the second term.  So now change variables to $\xi = \omega_p- \mu$, then  the renormalized gap equation is now:
\ba
1&&=  \frac{\kappa_R^2}{\pi^2}   \int_{\xi_{min}} ^ {\xi_{\max}}  d \xi  ~(\xi+\mu) \sqrt{ (\xi+\mu)^2 - m^2} \left[ \frac{1}{\sqrt{\Delta^2+
\xi^2}} + \frac{1}{\sqrt{\Delta^2+ (2 \mu+\xi)^2 } }   -\frac{2}{\xi+\mu}     \right]  \nonumber \\
&&=\frac{\kappa_R^2}{\pi^2}  \int_{\xi_{min}} ^ {\xi_{\max}}  d \xi ~ h_1(\xi,\mu,m, M) 
\equiv \frac{\kappa_R^2}{\pi^2}  f(\mu,m,m_H ,M).
\ea

Here  $\epsilon_{max} = \sqrt{m_H^2+m^2} - \mu \approx m_H$ and $\epsilon_{\min} = m- \mu$.

\begin{figure} 
   \includegraphics[width=3.4in ]{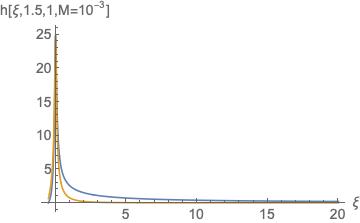}
   \caption{\label{hcompare}  Integrands  as a function of $\xi $ for$ \mu=1.5, m=1, m_, M=10^{-3}, $  $h_1$ is blue and $h_2$ is red}
\end{figure}

Now at large $\xi$ 
\bq
 h_1(\xi,\mu,m, M)  \rightarrow  -\frac{-2 \mu ^2+m^2+4 M}{\xi}.
 \eq
There is a logarithmic tail to $f$   coming  from this $1/\xi$ behavior as well as a large contribution coming from the sharp peak near the Fermi surface where $\omega_p  \approx \mu$ or $\xi \approx 0$.
 This is seen in our plot of $h_1[\xi] $ for $m=1,\mu=1.5, M=10^{-3}$ shown in Fig. (\ref{hcompare} ).

\begin{figure} 
   \includegraphics[width=3.4in ]{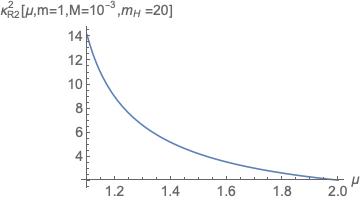}
   \caption{\label{ksqapprox} $\kappa_{Rpole} ^2$  as a function of $\mu $ for,$ m=1, m_, M=10^{-3},  m_H=20$ .}
\end{figure}

Now that we have a simple equation  for the renormalized gap eqauation at $\theta^*$ =0,
we can get an approximate analytic answer for the gap equation when $\Delta << m$ by approximating the $p^2$ term in the integrand
at the Fermi surface  $\omega_p=\sqrt{p^2+m^2} = \pm \mu$. We will be ignoring logarithmic corrections to the integrand coming from the tail. 
  If we replace the prefactor by its value at $\xi=0$ we get the approximation

\ba
1&&=  \frac{\kappa_R^2}{\pi^2}   \int_{\xi_{min}} ^ {\xi_{\max}}  d \xi  ~\mu \sqrt{ \mu^2 - m^2} \left[ \frac{1}{\sqrt{\Delta^2+
\xi^2}} + \frac{1}{\sqrt{\Delta^2+ (2 \mu+\xi)^2 } } -\frac{2}{\xi+\mu}    \right]  \nonumber \\
&&=\frac{\kappa_R^2}{\pi^2}  \int_{\xi_{min}} ^ {\xi_{\max}}  d \xi  h_2(\xi,\mu,m, M)
\equiv  \frac{\kappa_R^2} {\pi^2}  g_2(\mu,m, M,H) ,
\ea
where we have used the shorthand $H=m_H$.  We can evaluate $g_2$ analytically to obtain
\ba
&&g_2(\mu,m, M,H)= \mu \sqrt{\mu^2-m^2} \times \nonumber \\
   &&\left(\log \left(\sqrt{4 M+y^2}+y\right)+\log \left(2 \mu +\sqrt{4 M+(2 \mu +y)^2}+y\right)-2 
   \log (\mu +y) \right) |_{y={\xi_{\min}}}^{y={\xi_{\max}}} . \nonumber \\
   \ea
The integrands $h_1$ and $h_2$ are quite similar except for the $1/\xi$ tail in $h_1$.
This can be seen in Fig. (\ref{hcompare} )
This leads to the result that $g_1 > g_2$.  In terms of the value of $\kappa_R^2$, keeping only the pole contribution  and using the equation
\bq
\kappa^2_{R pole} = \frac{\pi^2}{g_2(\mu,m, M,H)}, 
\eq
we get the result shown in Fig. \ref {ksqapprox}.  The result of not including the logarithmic tail is that the approximate  value of $\kappa_{Rpole} ^2$ overestimates $\kappa_R^2$. This is seen by comparing   Fig. \ref {ksqapprox} with Fig.  \ref{ksqcompare}.  

As we let $\Delta^2$ go to zero the pole contribution gets larger and larger and the pole contribution becomes more important than the logarithmic tail.  

To get the usual type gap equation for $\Delta^2$ we can make some further approximations. 
 \bq
 \log \left(\sqrt{\Delta ^2+{\xi_{max}} ^2}+\xi _{max} \right)   \rightarrow \log{2 \xi_{max}} .
 \eq
  In the second $\log$  one has that $\xi_{min}  <0$ so when we  expanding the $\log$  for small $\Delta$ we obtain
\bq
\left(\sqrt{\xi_{min}^2}+\xi_{min}\right)+\frac{\Delta ^2}{2
   \sqrt{(\xi_{min})^2}}+O\left(\Delta ^3\right)  \approx \frac{\Delta ^2}{2
   \sqrt{(\xi_{min})^2}}.
\eq 
So we approximately get:
\bq
 \log \left(\sqrt{\Delta ^2+(\xi_{min})^2 } +\xi _{min}\right)   \rightarrow \frac{\Delta ^2}{2
   \sqrt{{\xi_{min}}^2}}.
   \eq
\ba
1&&= \frac{\kappa^2}{\pi^2} \mu  \sqrt{\mu ^2-m^2}  \log[ \frac{ 4\xi_{\max} \sqrt{\xi_{min}^2}}{\Delta^2}] .
\ea
This leads to the approximate equation
\bq
 \Delta^2 = 4 m_H (\mu -m) e^{-\frac{\pi ^2}{\kappa_R ^2 \mu  \sqrt{\mu ^2-m^2}}} ,
\eq
since $\xi_{max}  \approx m_H$ and $\xi_{min} = m- \mu <0$.   If we now consider the constraints on $\kappa_R^2 \approx G_{eff} $ coming from
the recent review of \cite{Review}  the maximum value of $ \rm{ log_{10} } ( \rm{G_{eff} MeV^2}) = -1.72$ . If we now a value of $ m= 1/2 ev$  and $m_H = 3 \times 10^4 ev$ this would lead to a negligible value of $\Delta^2$.

\section{Conclusions}
We considered a simple model for the  neutrino mass based on having a separate Higgs particle of mass 
$m_H$ coupling to one species of Dirac neutrinos.  By introducing composite fields connected with the two attractive channels we obtain  the effective potential for the composite fields  by making a Hubbard-Stratonovich transformation and integrating out the underlying fermion fields. We then keep the leading order term in the loop expansion of the resulting path integral expressed in terms of the composite fields  \cite{AF}. 
At finite density we find that there are two possible condensates having different quantum numbers,  coming from this interaction when viewed in the $``s"$ channel, having a mixing angle $\theta$.   What we find is that the theory
favors a very small mixing angle.
The second relative minimum solution, which has the condensate at the endpoint solution made only of the second condensate,  is very sensitive to the values of the coupling as well as $\mu$.  This solution always has higher energy than the true minimum. The predominant condensate is  of the form (in two component notation)  $(\nu_R \sigma_2 \nu_R) + (\nu_L \sigma_2 \nu_L)$.  

In our calculation, all the renormalized parameters such as the renormalized coupling constant and the renormalized effective potential are functions of $m, m_H, \mu, \Delta^2$  so in our plots it was easiest fix $\Delta^2$ at the value $10^{-3}$ to see how these functions depended on
$\mu$ and $m_H$. .  For this choice of $\Delta^2$ the effective coupling   $\rm{G_{eff} (eV)^2} $ had to be of order unity, which amounts to saying that $ \rm{ log_{10} } ( \rm{G_{eff} MeV^2}) ~12$. However, as discussed in the introduction, this is about 14 orders of magnitude larger than the value allowed by cosmological constraints.  The 31 orders of magnitude enhancement compared to the conventional Higgs mechanism was not enough. We conclude that for realistic values of the parameters, the gap $\Delta^2$ remains  exponentially small. 

Related work can be found in references \cite{Wang}   and \cite{Azam}. 

\begin{appendix}
\section{ \label{Fierz} Fierz Transformations}
The general Fierz transformation is based on the fact that there are sixteen independent $4 \times  4$ matrices  ${O}^i$ which can be written in terms of  five types of terms:  scalar $1$ vector  $\gamma^\mu$, tensor $\sigma^{\mu \nu}$, psudoscalar  $\gamma^5$ and axial vector $ \gamma^5 \gamma^\mu$.  So  we can write
\bq
\bar{\psi}_{1a}  M_{ab} \psi_{2b}\bar{\psi}_{3c} N_{cd} \psi_{4d}= \bar{\psi}_{1a} \bar{\psi}_{3c} \psi_{4d}\psi_{2b} M_{ab}  N_{cd} .
\eq
We can think the term $ M_{ab}  N_{cd}$   as the $ac$ component of a $4 \times 4$ matrix:
\bq
M_{ab}  N_{cd} = [K_{db}]_{ac} = C^i_{db} [O_i]_{ac} ,
\eq 
where $O_i = (O^i) ^{-1}$.  Then writing $ C^i_{db}=C^i_jO^j_{db}$  and using
\bq
Tr [O^j O_k] = 4 \delta^j_k ,
\eq
 we obtain:
\bq
( \bar{\psi} \psi )^2 =- \frac{1}{4} \sum_{\alpha=1}^6  \eta_\alpha (\psi^\dag O^\alpha \psi^\dag )(\psi O^{ \alpha \star} \psi) ,
\eq
and we have used the fact that only the antisymmetric matrices of the sixteen  $O^i$  can contribute since the $\psi_a$ anticommute. Here $\eta_\alpha= \pm 1$.  One finds  for  the six  non-vanishing $O^\alpha$  
\ba
\eta_\alpha&& =-1 ~~{\rm for} ~ ~\gamma^1, \gamma^3, \gamma^0 \gamma^5, \sigma^{02} \nonumber \\
\eta_\alpha&& =+1~~ {\rm for} ~ ~ \gamma^2 \gamma^5, \sigma^{13}  .\nonumber \\
\ea
Here  we have ($\mu \neq \nu$) 
\bq
\sigma^{\mu \nu} =  i  \gamma^\mu \gamma ^\nu .
\eq

\end{appendix}

\end{document}